\pdfoutput=1

\documentclass[11pt]{article}

\usepackage[final]{acl}

\usepackage{times}
\usepackage{latexsym}
\usepackage{amsmath}
\usepackage{amssymb}
\usepackage{booktabs}  
\usepackage{multirow}
\usepackage{adjustbox}
\usepackage{enumitem}
\usepackage[T1]{fontenc}
\usepackage{xcolor} 
\usepackage[utf8]{inputenc}

\usepackage{microtype}

\usepackage{inconsolata}

\usepackage{graphicx}
\usepackage{amsmath}
\usepackage{cuted}
\usepackage{xspace}
\newcommand{\method}{RecPO\xspace}
\renewcommand{\thefootnote}{\arabic{footnote}}
\usepackage{hyperref}
\usepackage{hyperref}
\definecolor{refcolor}{HTML}{00693e}
\definecolor{custompurple}{RGB}{165,119,233}
\definecolor{dartgreen}{HTML}{00693e}

\hypersetup{
    colorlinks=true,
    linkcolor=refcolor,
    citecolor=refcolor,
    filecolor=magenta,      
    urlcolor=refcolor,
    }

\title{What Makes LLMs Effective Sequential Recommenders?\\A Study on Preference Intensity and Temporal Context}

\author{
  Zhongyu Ouyang$^{1*}$ \quad Qianlong Wen$^{2*\ddagger}$ \quad Chunhui Zhang$^1$ \\
  \textbf{Yanfang Ye}$^3$ \quad \textbf{Soroush Vosoughi}$^{1\dagger}$ \\
  $^1$Department of Computer Science, Dartmouth College \\
  $^2$ ByteDance, US \\
  $^3$Department of Computer Science and Engineering, University of Notre Dame \\
  \small{\texttt{\{zhongyu.ouyang.gr, chunhui.zhang.gr, soroush.vosoughi\}@dartmouth.edu}} \\
  \small{\texttt{qianlong.wen@bytedance.com, yye7@nd.edu}}
}

\begin{document}
\maketitle

\begingroup
\renewcommand{\thefootnote}{}
\footnotetext{$^*$Equal contribution. $^\dagger$Corresponding author.}
\footnotetext{$^\ddagger$ Work done during PhD at University of Notre Dame.}
\endgroup

\begin{abstract}
What enables large language models (LLMs) to effectively model user preferences in sequential recommendation?
Our investigation reveals that existing preference-alignment approaches largely rely on binary pairwise comparisons, overlooking two critical factors: \textit{preference intensity}---the structured strength of affinity or aversion ---and \textit{temporal context}---the extent to which recent interactions better reflect a user's current intent.
Through controlled experiments, we show that leveraging comprehensive feedback with structured preference signals substantially improves recommendation performance, indicating that binary modeling discards essential information.
Motivated by these findings, we propose \method, a unified preference optimization framework that maps both explicit and implicit feedback into a common preference signal and constructs adaptive reward margins that jointly account for preference intensity and interaction recency.
Experiments across five datasets show that \method consistently outperforms state-of-the-art baselines while exhibiting behavioral patterns aligned with human decision-making, including favoring immediate satisfaction, maintaining preference coherence, and avoiding dispreferred items.
Our results highlight that preference intensity and temporal context are fundamental ingredients for effective LLM-based recommendation.
Code: \url{https://github.com/zyouyang/RecPO}
\end{abstract}

\section{Introduction}\label{sec:intro}
Large language models (LLMs) are increasingly being adapted for sequential recommendation~\cite{harte2023leveraging,li2023text,yang2024sequential,bao2023tallrec,zhang2023recommendation}, where the task is to predict the next item a user will interact with based on their historical behaviors.
Unlike traditional recommenders~\cite{hidasi2016session,kang2018self,tang2018personalized}, LLM-based systems~\cite{chen2024elcorec} leverage semantic understanding and reasoning capabilities to model user preferences from textual interaction histories from multiple perspectives.

\begin{figure}[t]
  \centering
  \includegraphics[width=0.9\linewidth]{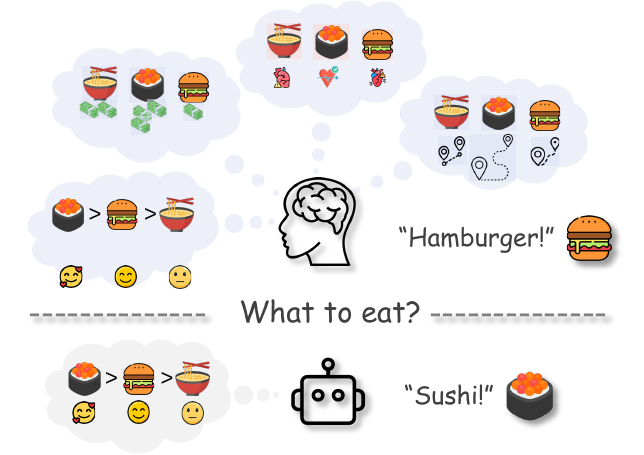}
  \caption{Human decision-making involves trade-offs among \textit{preference intensity, temporal context, effort, and risk}---factors that are largely overlooked in current LLM-based preference modeling.}
  \label{fig:intro}
\end{figure}

Current approaches predominantly rely on preference alignment techniques such as DPO~\cite{rafailov2024direct} and its variants~\cite{chen2024on,meng2024simpo,amini2024direct}, which treat all preferences uniformly through binary pairwise comparisons.
This binary abstraction, while effective for general language tasks, misaligns with human decision-making: humans exhibit structured preferences (\textit{strongly} love vs. \textit{mildly} like) and recency-sensitive temporal preferences, where more recent interactions better reflect current intent than older ones, as illustrated in Figure~\ref{fig:intro}.
Such structured and temporally contextualized patterns are pervasive in human behavior~\cite{astington1995theory}, yet remain unmodeled in current LLM-based recommenders.
This raises a critical question: \textit{what specific factors in preference data enable LLMs to capture these nuanced human behaviors for recommendation?}

We investigate this question through systematic empirical study.
Our proof-of-concept experiment (\S~\ref{sec:proof_exp}) reveals that incorporating \textit{comprehensive feedback} (including negative interactions) and \textit{structured preference signals} (e.g., ratings) substantially improves performance, indicating that binary modeling discards critical information.
Through controlled ablations, we identify two key factors: \textbf{(1) preference intensity}---the structured strength of user affinity or aversion, and \textbf{(2) temporal context}---the extent to which more recent interactions better reflect a user’s current intent.
These factors, though well-established in behavioral economics and cognitive science, have been largely overlooked in LLM preference alignment for recommendation.

Building on these insights, we introduce \method, a preference optimization framework that operationalizes preference intensity and temporal context via adaptive reward margins.
Unlike prior work that applies uniform margins across all preference pairs, \method leveraged fine-grained preference signals---\textit{(i)} structured preference strength (e.g., 5-star vs. 3-star ratings), and \textit{(ii)} interaction recency relative to the current decision point---to modulate preference alignment.
As a result, \method enables LLMs to model evolving user preferences in a manner more consistent with human decision-making.
Our contributions are threefold:
\begin{itemize}[leftmargin=*]
    \item We systematically demonstrate that preference intensity and recency-sensitive temporal context are critical factors for LLM-based preference modeling in sequential recommendation(\S~\ref{sec:proof_exp}).
    \item We propose \method, a unified preference optimization framework that incorporates these factors via adaptive reward margins, enabling effective preference alignment for sequential recommendation (\S~\ref{sec:method}).
    \item Through experiments on five datasets with both explicit and implicit feedback, we show that \method improves recommendation accuracy and exhibits human-aligned behaviors by prioritizing items aligned with current user intent and maintaining coherent preferences under shifting contexts (\S~\ref{sec:exp}).
\end{itemize}

\section{Preliminaries}
\label{sec: preliminary}
\paragraph{Sequential Recommendation with LMs.}
We begin by formalizing the sequential recommendation task within the LM framework. 
Let $\mathcal{H}_u=[i^1, i^2, ..., i^{N_u}]$ represent the chronologically ordered sequence of historical interactions for user $u$, where each element $i^k$ encapsulates contextual details of the $k$-th interaction (e.g., item title, style, rating), and $N_u$ denotes the total number of interactions.
We define $\mathcal{H}_u^t=\mathcal{H}_u[\colon t]$ as as the subset of interactions up to time $t$, and let $i_p^{t+}$ denote the \textit{next recent favorable (highly-rated)} item following the interaction history at $t$.
Let $\mathcal{\pi}_\theta$ be the LM performing the task, parameterized by $\theta$.
The sequential recommendation task within the LM framework is formulated as follows:
given user $u$'s interaction history $\mathcal{H}_u^t$ up to time $t$ and a candidate item set $\mathcal{C} = \{i^{(j)}\}_{j=1}^{K}$, where $\mathcal{H}_u^t \cap \mathcal{C} = \emptyset$ and $i_p^{t+} \in \mathcal{C}$, the model $\pi_\theta$ is required to predict the item that most likely be favorable to user, i.e., $i_p^{t+}$. 

\paragraph{Adapting and Aligning LMs with Human Preference Feedback.}
Existing LMs are adapted to sequential recommendation tasks through a two-stage training paradigm, namely \textit{supervised fine-tuning (SFT)}~\cite{ouyang2022training,llara2024liao,bao2023tallrec}, which adapts general-purpose LLMs into task-specific models, and \textit{preference alignment}~\cite{schulman2017proximal,ouyang2022training}, which further aligns model output to human preference\footnote{More detailed preliminaries in Appendix~\ref{appd:prelim}.}.

In \textit{SFT}, the model is trained to predict the target item given a user's interaction history and contextual information.
Let $\mathbf{x}_u^t$ be the task prompt constructed from user $u$’s interactions up to time $t$, and let $\mathbf{y}_p^t$ be the textual mapping of the target item. The SFT objective is:
\begin{equation}\label{eq:sft}
    \min_\theta \quad -\mathbb{E}_{(\mathbf{x}_u^t, \mathbf{y}^{t}_p)\sim \mathcal{D}_{\text{SFT}}} \left[ \log \pi_\theta(\mathbf{y}^t_p|\mathbf{x}_u^t)  \right].
\end{equation}
The resulting model is denoted as $\pi_{\text{SFT}}$. For brevity, we omit time superscripts when unambiguous.

While SFT adapts LMs to the task format, recent studies indicate that models still struggle to align outputs with human judgments of quality~\cite{ziegler2019fine,stiennon2020learning,rafailov2024direct}.
To address this, preference alignment further optimizes models using preference data.
A representative method is DPO~\cite{rafailov2024direct}, which models pairwise preferences using the Bradley–Terry framework~\cite{bradley1952rank}.
Given a preferred–dispreferred output pair $(\mathbf{y}_p, \mathbf{y}_d)$, the DPO objective is:
\begin{align}\label{eq:dpo}
    \min_\theta & -\mathbb{E}_{(\mathbf{x}, \mathbf{y}_p, \mathbf{y}_d) \sim D} \Bigl[ \log \sigma \Bigl( \beta \log \frac{\pi_\theta (\mathbf{y}_p | \mathbf{x})}{\pi_{\text{ref}}(\mathbf{y}_p | \mathbf{x})} \notag\\
    & - \beta \log \frac{\pi_\theta (\mathbf{y}_d | \mathbf{x})}{\pi_{\text{ref}}(\mathbf{y}_d | \mathbf{x})} \Bigr) \Bigr],
\end{align}
where $\pi_{\text{ref}}$ is typically set to $\pi_{\text{SFT}}$ and $\beta$ controls the strength of preference alignment.

For sequential recommendation, S-DPO~\cite{chen2024on} extends DPO by pairing each preferred item with multiple dispreferred items $\mathcal{T}_d$, yielding the objective:

{
\fontsize{10.5pt}{12pt}\selectfont
\begin{align}\label{eq:s-dpo}
    \min_\theta &-\mathbb{E}_{(\mathbf{x}, \mathbf{y}_p, \mathcal{T}_d) \sim D} \biggl[ \log \sigma \biggl( -\log \sum_{\mathbf{y}_d \in \mathcal{T}_d} \exp \Bigl( \notag \\
     & \beta \log \frac{\pi_\theta (\mathbf{y}_d | \mathbf{x})}{\pi_{\text{ref}}(\mathbf{y}_d | \mathbf{x})} - \beta \log \frac{\pi_\theta (\mathbf{y}_p | \mathbf{x})}{\pi_{\text{ref}}(\mathbf{y}_p | \mathbf{x})} \Bigr) \biggr) \biggr],
\end{align}
}

\noindent
where $\mathcal{T}_d$ denotes the set of dispreferred items\footnote{We use positive/negative, as well as preferred/dispreferred interchangeably in the following content.}.
\begin{figure}[t]
  \centering
  \includegraphics[width=\linewidth]{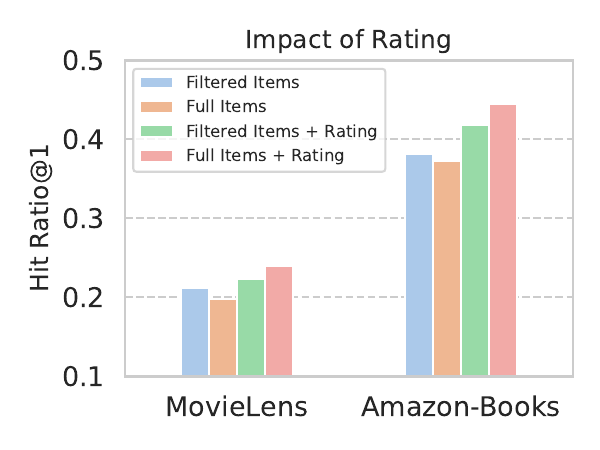}
  \caption{Hit@1 in next favorable item prediction with comprehensive and structured preference feedback.
  }
  \label{fig:motivation}
\end{figure}

\begin{figure*}[ht]
  \centering
  \includegraphics[width=\linewidth]{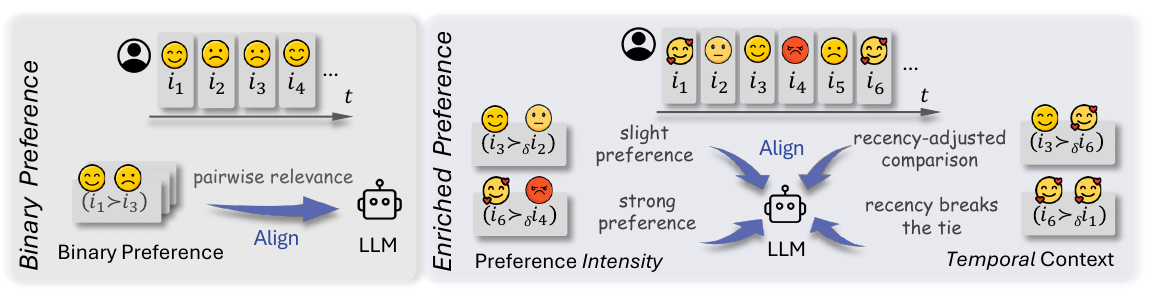}
  \caption{Illustrations for preference learning frameworks with binary and enriched preference: the prior assumes binary distinction in preference, while the latter enriches preference distinction with preference intensity and temporal context  ($\delta$ indicates the enrichment).}
  \label{fig:main}
\end{figure*}

\section{What Do Current Methods Overlook? A Proof-of-Concept Investigation}
\label{sec:proof_exp}
Current LLM-based recommenders, including S-DPO~\cite{chen2024on}, typically filter out negative feedback items from user histories and discard structured preference signals, treating all remaining items uniformly.
But does this practice discard critical information?

To investigate this, we design a proof-of-concept experiment that varies two dimensions of user feedback: \textit{comprehensiveness} (whether negative interactions are retained) and \textit{structure} (whether structured preference signals are provided). 
We consider four input configurations: 
(i) \textit{Filtered Items}: excluding negative interactions, without explicit ratings (mimicking S-DPO’s setup);
(ii) \textit{Full Items}: retaining all interactions, without ratings;
(iii) \textit{Filtered Items + Rating}: excluding negative interactions, with ratings;
(iv) \textit{Full Items + Rating}: retaining all interactions with corresponding ratings.

Using SFT only, we fine-tune LLaMA3-8B on MovieLens and Amazon-Books (described \S~\ref{sec:setup}) under each configuration.
Performance is evaluated using Hit Ratio@1 (see \S~\ref{sec:setup}, higher is better), with results shown in Figure~\ref{fig:motivation}. 

\paragraph{Key Findings.}
\textbf{(1) Comprehensive feedback enables aversion modeling.} Comparing \textit{Filtered Items + Rating} vs. \textit{Full Items + Rating}, retaining negative interactions consistently improves performance.
This indicates that negative interactions provide informative aversion signals that help delineate user preferences when their strength is explicitly encoded.
\textbf{(2) Structured signals are required to recover preference intensity.} Comparing \textit{Full Items} vs. \textit{Full Items + Rating}, adding structured preference annotations yields substantial gains.
Without such signals, \textit{Full Items} underperforms \textit{Filtered Items}, as unannotated negative interactions collapse into noise.
\textbf{(3) Both factors are jointly necessary.} The best performance is achieved only when comprehensive feedback is combined with structured preference signals, suggesting that effective preference modeling requires access to both the full interaction spectrum and structured preference information.

These results indicate that existing methods overlook the aspects of structured strength of preferences (preference intensity) and the informative role of negative interactions in delineating user aversion.
A natural question then arises: how can LLMs effectively exploit such fine-grained signals when they are embedded within a user’s interaction sequence?
Crucially, when preferences are modeled as structured signals over time, their contribution to future decisions becomes nonstationary.
Interactions with similar preference intensity may differ substantially in relevance depending on their temporal proximity to the current decision.
This observation motivates our study of \textbf{preference intensity} and \textbf{recency-sensitive temporal context} as complementary factors for enabling LLMs to capture richer, human-aligned preference dynamics.

\section{Methodology}
\label{sec:method}
We first lay out the prompt design underlying preference modeling in LLM-based recommendation, and then present \method, a preference optimization framework that calibrates reward margins using structured and temporal contextualized preference feedback (Figure~\ref{fig:main}).

\subsection{Operationalizing Comprehensive and Structured Feedback}
Instead of discarding negatively interacted items to construct homogeneous histories~\cite{llara2024liao, chen2024on}, we retain each user’s complete interaction sequence, including both positive and negative feedback.
Each historical item is paired with an associated preference signal, obtained either from explicit ratings or from implicit feedback converted into structured scores, yielding a structured preference profile.
Following prior work~\cite{chen2024on}, the input prompt is composed of the following components:

\paragraph{User historical interaction $\mathcal{H}_u$}
Each item in the user history is formatted as "\texttt{[ItemTitle] | Rating: [ItemRating]}".
For example, "\texttt{Toy Story | Rating: 4}".
All historical items are concatenated with "\texttt{\textbackslash n}" being the separator.

\paragraph{Candidate item set $\mathcal{C}$}
We format all candidate items in a similar format as the historical items, except that no preference attributes are provided.

\paragraph{Task Description}
We prepend the history-specific prefixes (e.g., "\texttt{Given the user’s recent viewing and rating history}") and candidate-specific prefixes (e.g., "\texttt{recommend a movie they will likely watch next and rate generously from following candidates}") to their respective sequences.
The three prompt components are concatenated as the final textual input $\mathbf{x}_u$ to the LMs.
Concrete examples are demonstrated in Appendix~\ref{appd:prompt}.

\subsection{Modeling \textit{Preference Intensity} and \textit{Temporal Context}}
We extend standard pairwise preference optimization by introducing an adaptive reward margin $\gamma_r$ that is determined by both the structured preference strength of the compared items and their relative recency with respect to the current decision point.
\begin{figure*}[t]
\centering
\setcounter{equation}{6}
\begin{align}
\mathcal{L} \left(\pi_\theta ; \pi_{\mathrm{ref}}\right) & = -\mathbb{E}_{(\mathbf{x}_u, \mathbf{y}_p, \mathcal{T}_d) \sim \mathcal{D}} 
 \Biggl[ \log \sigma \Biggl( \nonumber -\log \sum_{\mathbf{y}_d \in \mathcal{T}_d} \exp \Biggl( \beta \log \frac{\pi_\theta(\mathbf{y}_d \mid \mathbf{x}_u)}
{\pi_{\mathrm{ref}}(\mathbf{y}_d \mid \mathbf{x}_u)} \nonumber\\
&\quad - \beta \log \frac{\pi_\theta(\mathbf{y}_p \mid \mathbf{x}_u)}
{\pi_{\mathrm{ref}}(\mathbf{y}_p \mid \mathbf{x}_u)} - \lambda \frac{\phi(s_p, \Delta_{t_p})}
{\phi(s_d, \Delta_{t_d})} \Biggr) \Biggr) \Biggr].
\label{eq:loss}
\end{align}
\end{figure*}
Specifically, we define $\gamma_r$ for a preference pair $(\mathbf{y}_p, \mathbf{y}_d)$ as:
\setcounter{equation}{3}             
\refstepcounter{equation}            

\begin{equation}
\tag{\theequation}                   
\label{eq: margin}                    
\begin{aligned}
\gamma_r = \lambda  \frac{\phi\left(s_p, \Delta_{t_p}\right)}{\phi\left(s_d, \Delta_{t_d}\right)}
\end{aligned}
\end{equation}
where $\mathbf{y}_p$ is preferred over $\mathbf{y}_d$, $\Delta_{t_p}=t_p^{+}-t$ represents the temporal distance between the interaction and the current decision point, and $\lambda$ controls the margin magnitude.
The terms $s_p$ and $s_d$ are their structured preference scores derived from explicit or implicit feedback; they are defined abstractly and may incorporate multiple forms of preference evidence rather than being tied to a single signal.
The utility function $\phi(\cdot)$ \textit{jointly} captures preference intensity and temporal relevance.

In this work, we instantiate $\phi\left(s, \Delta_{t}\right) = s / \left(\Delta_{t}\right)^{0.5}$, though alternative forms of the general family $\phi(s, \Delta_t) \propto s / (\Delta_t)^{\alpha}$ with $\alpha > 0$ are also compatible.
For dispreferred items from negative sampling or historical interactions without explicit user feedback, we assign default preference scores and time latencies to facilitate training.
More implementation details are provided in \S~\ref{sec:exp}.

\subsection{Deriving the Preference Alignment Objective}
We plug Equation~\ref{eq: margin} into the BT model to derive the distribution for pairwise preference data:
\begin{equation}
\begin{aligned}
P^*(\mathbf{y}_p \succ & \mathbf{y}_d  \mid \mathbf{x}_u) = \\
& \sigma\left(
r\left(\mathbf{x}_u, \mathbf{y}_p\right) - r\left(\mathbf{x}_u, \mathbf{y}_d\right) - \gamma_r
\right),
\end{aligned}
\label{eq:BT}
\end{equation}
where $r(\cdot)$ is the reward function, and $\sigma(\cdot)$ is the sigmoid function.
We pair each preferred item with multiple dispreferred items, and leverage the Plackett-Luce (PL) model~\cite{pl1, pl2} to generalize pairwise comparisons to a list-wise ranking framework. 
Formally, given the prompt $x_u^t$ encompassing all the historical interactions of user $u$, a candidate set $\mathcal{C}$ containing $K$ items (one preferred item and $K-1$ dispreferred items), and a permutation $\tau$ representing the predicted ranking of these candidates based on user preference for the next item (denote $\tau(j)$ as the item ranked at position $j$), the probability of observing the candidates' preference ranked as $[ \mathbf{y}_{\tau(1)}, \mathbf{y}_{\tau(2)}, \ldots, \mathbf{y}_{\tau(K)} ]$ is:
\begin{align}
\label{eq:PL}
P(\tau \mid &\mathbf{x}_u, \mathcal{T}_c)= \nonumber \\
&\prod_{j=1}^K \frac{\exp \left(r\left(\mathbf{x}_u, \mathbf{y}_{\tau(j)}\right)\right)}{\sum_{m=j}^K \exp \left(r\left(\mathbf{x}_u, \mathbf{y}_{\tau(m)}\right)\right)},
\end{align}
where $\mathcal{T}_c$ contains $K$ item descriptions. 
Building upon Equation~\ref{eq:PL}, we derive the final objective shown in Equation~\ref{eq:loss}.
Note that our method is reduced to S-DPO when $\lambda=0$. 
For brevity, the detailed derivation process is provided in Appendix~\ref{appd:derivation}. 
Optimizing the derived objective effectively integrates structured preference intensity with recency-sensitive temporal context, enabling LLM recommenders to align preference learning with users’ evolving intent in sequential decision settings.
\begin{table*}[t]
\begin{adjustbox}{width=\textwidth,center}
\begin{tabular}{c|c|c|cc|cc|cc|cc|cc}
\toprule
\multirow{2}{*}{\shortstack{\textbf{Model}\\\textbf{Type}}} & \multirow{2}{*}{\textbf{Bkbn}} & \multirow{2}{*}{\textbf{Method}} & \multicolumn{2}{c|}{\textbf{MovieLens}} & \multicolumn{2}{c|}{\textbf{Amazon-Books}} & \multicolumn{2}{c|}{\textbf{BeerAdvocate}} & \multicolumn{2}{c|}{\textbf{Steam}} & \multicolumn{2}{c}{\textbf{LastFM}} \\
& & & HR@1 & ValidRatio & HR@1 & ValidRatio & HR@1 & ValidRatio & HR@1 & ValidRatio & HR@1 & ValidRatio \\
\midrule
\multicolumn{3}{c|}{Feedback Type} & \multicolumn{6}{c|}{Explicit Feedback} & \multicolumn{4}{c}{Implicit Feedback} \\
\midrule
\multirow{3}{*}{\rotatebox{90}{Trad.}} 
  & - & GRU4Rec  & 0.2664 & 1.0000 & 0.1310 & 1.0000 & 0.3708 & 1.0000 & 0.4584 & 1.0000 & 0.6630 & 1.0000 \\
  & - & Caser    & 0.2714 & 1.0000 & 0.1538 & 1.0000 & 0.3757 & 1.0000 & 0.4394 & 1.0000 & \underline{0.6716} & 1.0000 \\
  & - & SASRec   & 0.2671 & 1.0000 & 0.1559 & 1.0000 & 0.3800 & 1.0000 & \underline{0.4587} & 1.0000 & 0.6659 & 1.0000 \\
\midrule
\multirow{11}{*}{\rotatebox{90}{LLM}} 
  & \multirow{6}{*}{\rotatebox{90}{LLaMA3-8B}} 
    & LLaMA3     & 0.0929 & 0.7351 & 0.0654 & 0.6165 & 0.0686 & 0.6617 & 0.0852 & 0.8672 & 0.1264 & 0.6147 \\
  &  & SFT        & 0.2478 & 0.9985 & 0.4447 & 0.9974 & 0.2645 & 0.9936 & 0.3122 & 0.9990 & 0.5076 & 1.0000 \\
  &  & DPO        & 0.2809 & 0.9970 & 0.5049 & 0.9887 & 0.4412 & 0.9875 & 0.3340 & 0.9980 & 0.5719 & 1.0000 \\
  &  & SimPO      & \underline{0.2974} & 0.9725 & \underline{0.5129} & 0.9564 & 0.4020 & 0.9250 & 0.3401 & 0.9766 & 0.5759 & 0.9419 \\
  &  & S-DPO      & 0.2902 & 0.9983 & 0.5065 & 0.9880 & \underline{0.4698} & 0.9903 & 0.3588 & 0.9990 & 0.5719 & 0.9990 \\
  &  & \method    & \textbf{0.3451} & 0.9969 & \textbf{0.5802} & 0.9851 & \textbf{0.5771} & 0.9887 & \textbf{0.4672} & 0.9985 & \textbf{0.6830} & 0.9959 \\
\cmidrule{2-13}
  & \multirow{6}{*}{\rotatebox{90}{Qwen-7B}} 
    & Qwen & 0.1204 & 0.7471 &   0.1013   & 0.7194   & 0.0583      & 0.4223    & 0.1477      & 0.6293       & 0.2148      & 0.6860      \\
  &  & SFT        & 0.2060 & 0.9983 & 0.3659 & 0.9967 & 0.2044 & 0.9849 & 0.2081 & 0.9950 & 0.3119 & 0.9969 \\
  &  & DPO        & 0.2610 & 0.9983 & 0.4412 & 0.9930 & 0.2600 & 0.9724 & 0.2457 & 0.9960 & 0.4046 & 0.9969 \\
  &  & SimPO      & 0.2888 & 0.9531 & 0.4644 & 0.9880 & 0.4044 & 0.9529 & 0.3706 & 0.9940 & 0.5209 & 0.9796 \\
  &  & S-DPO      & 0.2706 & 0.9957 & 0.4623 & 0.9910 & 0.3253 & 0.9798 & 0.3062 & 0.9970 & 0.4495 & 0.9959 \\
  &  & RecPO      & \textbf{0.3446} & 0.9896 & \textbf{0.5307} & 0.9880 & \textbf{0.4320} & 0.9729 & \textbf{0.4143} & 0.9912 & \textbf{0.5973} & 0.9980 \\
\bottomrule
\end{tabular}
\end{adjustbox}
\caption{Overall model performance comparison on five real-world recommendation datasets. The best performance is bolded, and runner-ups are underlined. Datasets are grouped by explicit and implicit feedback.}
\label{tab:overall}
\end{table*}

\section{Experiment}\label{sec:exp}
\subsection{Setup}\label{sec:setup}
\paragraph{Datasets.}
We use five widely used real-world sequential recommendation datasets for evaluation, including 
\textit{MovieLens-1M}~\cite{harper2015movielens}, 
\textit{Amazon-books}~\cite{ni2019justifying}, 
\textit{Steam}~\cite{kang2018self}, 
\textit{BeerAdvocate}~\cite{leskovec2012learning}, and \textit{LastFM}~\cite{celma2010music}.
More dataset details are provided in Appendix~\ref{appd:dataset}.

For each dataset, we apply $k$-core filtering~\cite{he2016vbpr} with $k=5$ to remove users and items with insufficient interactions.
We construct a candidate set of 20 items for next-item prediction.
During training, the candidate set includes 10 subsequent interactions (ensuring the ground-truth item is present) and 10 randomly sampled non-interacted items; during validation and testing, the set consists of the ground-truth item and 19 randomly sampled non-interacted items.
For MovieLens-1M, Amazon-Books, and BeerAdvocate, explicit ratings are used as structured preference signals, while for Steam and LastFM, where explicit ratings are unavailable, we use play-hours and play-count as proxies.
For each user, interactions are ordered chronologically, with the second-last interaction used for validation, the last for testing, and the remainder for training.

\paragraph{Baselines.}
We compare \method with two types of baseline models:
(i) \textit{Traditional} methods leverage sequential patterns in user behaviors to predict the next interacted item, using various modeling architectures such as recurrent neural networks (GRU4Rec~\cite{hidasi2016session}), convolutional neural networks (Caser~\cite{tang2018personalized}), or multi-head self-attention frameworks (SASRec~\cite{kang2018self});
(ii) \textit{LM-based} methods utilize LMs to process historical interactions and predict the next interacted item.
We select two LM backbones, LLaMA3~\cite{dubey2024llama} and Qwen~\cite{bai2023qwen}, and compare between the standard preference optimization baseline DPO~\cite{rafailov2024direct},
SimPO~\cite{meng2024simpo}, a reference-free method that enhances DPO with length regularization and a fixed margin term, and S-DPO~\cite{chen2024on}, which adapts DPO specifically for sequential recommendation.
More baseline baseline details are provided in Appendix~\ref{appd:baselines}. 

We \textit{exclude} proprietary LLMs due to their lack of training access in integrating preference intensity and temporal context.
As this work is hypothesis-driven rather than method-focused, our objective is to validate how these factors enable effective preference modeling, which requires full control over LLMs unavailable in closed-source systems.

\paragraph{Implementation.}
All experiments are performed on 8 NVIDIA RTX A100 with 80GiB of VRAM. 
For all the preference learning approaches, we first conduct SFT for task adaptation, and then post-train models initialized from SFT checkpoints by optimizing the alignment loss in Equation~\ref{eq:loss}\footnote{More implementation details in Appendix~\ref{appd:implementation}}.

\paragraph{Evaluation Metrics.}
We follow S-DPO and evaluate models using two metrics: (i) \textit{Hit Ratio@1} measures the proportion of test cases where the top-ranked item matches the ground-truth target, and (ii) \textit{Valid Ratio} captures instruction compliance by quantifying the fraction of outputs that follow formatting rules and remain within the candidate set. 
The latter ensures outputs are valid and in-distribution. 
Together, they assess both recommendation accuracy and practical deployability.

\subsection{Do Preference Intensity and Temporal Context Improve Recommendation?}
\label{sec:exp_overall}
\paragraph{Overall Performance.}
Table~\ref{tab:overall} compares \method with the baselines across the five datasets, revealing the following key findings:
\begin{itemize}[leftmargin=*]
    \item \textit{\textbf{SFT establishes base model capabilities.}} Raw LLMs, while possess rich world knowledge, frequently violate recommendation constraints. SFT substantially improves output validity, matching traditional recommenders and underscoring the need for task-specific adaptation. 
    \item \textit{\textbf{Binary preference modeling provides limited gains.}}
    Preference optimization methods consistently outperform SFT in Hit Ratio@1, but standard DPO provides modest improvements. Methods that incorporate multiple negatives (S-DPO and \method) perform better, while SimPO achieves higher accuracy at the cost of reduced output validity.
    This suggests that treating preferences as isolated binary comparisons underutilizes available preference information in sequential recommendation.
    \item \textit{\textbf{Integrating preference intensity and temporal context yields substantial improvements.}} 
    By integrating structured preference signals with recency-sensitive adaptive margins, \method consistently achieves the best performance across both LLM backbones. Gains over traditional recommenders are smaller on implicit-feedback datasets, likely due to the homogeneity of proxy-derived preference signals that even simple traditional models can effectively capture.
\end{itemize}

\paragraph{Do Preference Intensity and Temporal Context Both Matter?}
We ablate the contributions of preference intensity and recency-sensitive temporal context by progressively removing components of the adaptive margin. 
As shown in Table~\ref{tab:ablation_intensity_temporal}, removing the margin entirely (–I –T), which reduces RecPO to S-DPO, leads to consistent performance drops across datasets. 
Incorporating preference intensity alone (–T) substantially improves accuracy, confirming the importance of structured preference signals. 
\method achieves the best overall performance on most datasets, demonstrating that preference intensity and temporal context are complementary. 

\begin{table}[t]
\centering
\small
\begin{adjustbox}{width=\linewidth,center}
\begin{tabular}{l|ccccc}
\toprule
\multicolumn{6}{c}{\textbf{Hit Ratio@1}} \\
\midrule
Variant & ML & AmazonB & Beer & Steam & LastFM \\
\midrule
--I --T & 0.2902 & 0.5065 & 0.4698 & 0.3588 & 0.5719 \\
--T     & 0.3343 & 0.5661 & \textbf{0.6143}$^{*}$ & 0.4202 & 0.6544 \\
\method    & \textbf{0.3451} & \textbf{0.5802} & 0.5771 & \textbf{0.4672} & \textbf{0.6830} \\
\midrule
\multicolumn{6}{c}{\textbf{Valid Ratio}} \\
\midrule
Variant & ML & AmazonB & Beer & Steam & LastFM \\
\midrule
--I --T & 0.9983 & 0.9880 & 0.9903 & 0.9990 & 0.9990 \\
--T     & 0.9983 & 0.9798 & 0.9407$^{*}$ & 0.9980 & 0.9959 \\
\method    & 0.9969 & 0.9851 & 0.9887 & 0.9985 & 0.9959 \\
\bottomrule
\end{tabular}
\end{adjustbox}
\caption{Ablation study on preference intensity and recency-sensitive temporal context.
--I --T removes adaptive margins (equivalent to S-DPO);
--T uses preference intensity only;
$^{*}$\textit{Excessively low valid ratios are impractical for real-world deployment.}}
\label{tab:ablation_intensity_temporal}
\end{table}

\paragraph{How Should Preference Intensity and Temporal Context Be Combined?}
Let $\phi_p$ and $\phi_d$ be the scores for the preferred and dispreferred items respectively. 
By default, \method defines the margin term $\gamma_t$ as the ratio of preference scores $\phi$ between positive and negative item pairs (Equation~\ref{eq: margin}). 
To investigate how these factors should be mathematically combined, we introduce two alternative margin functions: \textit{(i) Log Diff}, $\gamma_r= \lambda \log\left(\phi_p-\phi_d\right)$; \textit{(ii) Log Ratio}, $\gamma_r= \lambda  \left(\log\phi_p-\log \phi_d\right)$. 
As shown in Table~\ref{tab:abl}, both variants outperform the strongest LLM-based recommender baseline, confirming the benefits of incorporating these factors through any margin formulation.
\method’s default ratio-based margin achieves the best overall performance by amplifying training gradients, especially when historical user ratings show low volatility. By directly contrasting $\phi_p$ and $\phi_d$ via division, stronger learning signals are provided to help the model prioritize subtle but critical preference patterns.
\begin{table}[t]
\begin{adjustbox}{width=0.9\linewidth,center}
\begin{tabular}{c|c|c|c}
\toprule
Dataset           & Log Diff  & Log Ratio & \method
\\ \midrule
MovieLens               &   0.3160        &     0.3247        &           0.3451        \\
Amazon-Books             &   0.5370 &  0.5455            &   0.5802                \\
BeerAdvocate &     0.5023      &     0.5257         &     0.5771 \\
Steam &   0.4284        &      0.4517        &   0.4672    \\
LastFM &     0.5912 &     0.6388      &   0.6830 
\\ \bottomrule   
\end{tabular}
\end{adjustbox}
\caption{Ablation study on the margin function, Hit Ratio@1 is reported for comparison.}
\label{tab:abl}
\end{table}

\begin{figure*}[t]
  \centering
  \includegraphics[width=\linewidth]{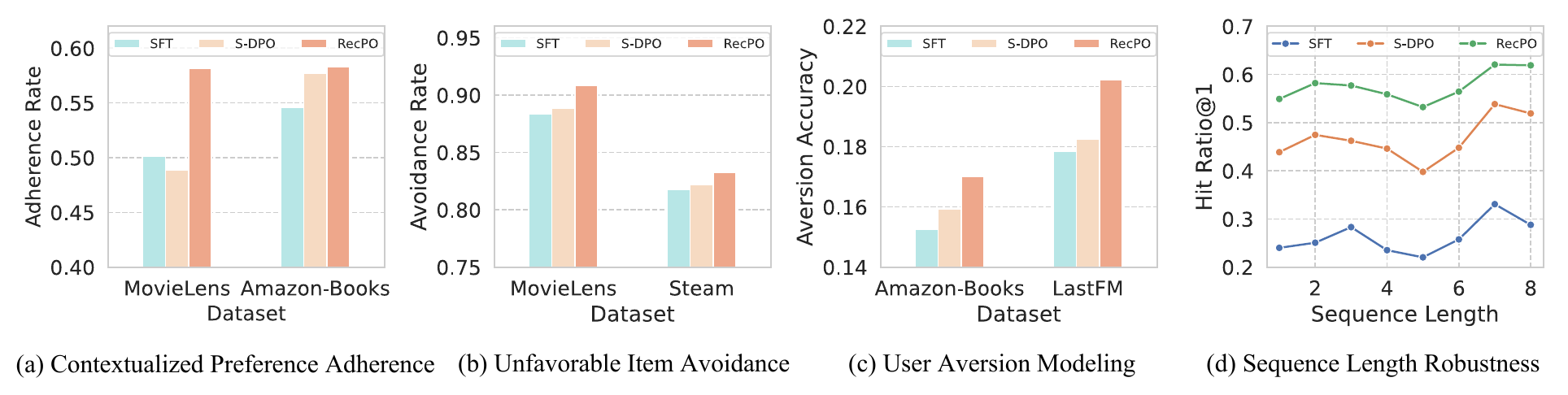}
  \caption{Comparing between SFT, S-DPO, and \method from the perspectives of adhering to contextual preference (a), avoiding unfavorable items under temptation (b), identifying dis-preferred items (c), and consistently performing across varying user history lengths (d). The adherence rate and avoidance rate are defined in \S~\ref{sec:exp_analysis}.}
  \label{fig:exp}
\end{figure*}
\subsection{Do Models Learn Human-Aligned Preference Patterns?}
\label{sec:exp_analysis}
Beyond quantitative performance, we investigate whether incorporating preference intensity and temporal context enables models to exhibit human-like decision patterns. We probe the learned preferences from multiple perspectives:
\begin{itemize}[leftmargin=*]
    \item \textbf{Temporal context sensitivity}: When the candidate set includes other future highly-rated items, does the model still prioritize the correct next item, reflecting sensitivity to temporal context?
    \item \textbf{Preference intensity awareness}: When the candidate set includes future low-rated items that may appear contextually tempting, can the model avoid recommending them?
    \item \textbf{Implicit aversion modeling}: When directly prompted, can the model correctly identify the item least aligned with the user’s preferences?
    \item \textbf{Robustness across contexts}: Does the model maintain stable performance across users with varying lengths of interaction history?
\end{itemize}

\paragraph{Temporal context enables immediate satisfaction prioritization.}
To assess \method’s ability to model contextualized preferences, we construct more challenging test sets for MovieLens and Amazon-Books by augmenting the candidate pool with other highly-rated items from users’ future interactions.
This setup tests whether the model can prioritize the correct next item when competing items, though eventually preferred, are not immediately relevant.
To this end, we define the \textit{Adherence Rate} as the proportion of cases in which the model recommends the next immediately preferred item.
A high Adherence Rate indicates that the model consistently recommends relevant and contextualized items among all highly-rated candidates.
A higher Avoidance Rate signifies that the model is better at resisting the temptations from the unfavorable items.
More details are included in Appendix~\ref{appd:metrics}.

As shown in Figure~\ref{fig:exp}(a), \method consistently outperforms both SFT and S-DPO, more reliably ranking the temporally appropriate item at the top.
This suggests improved sensitivity to short-term intent and temporal alignment.
In contrast, S-DPO fails to consistently outperform SFT, indicating a failure to fully capture context-dependent user goals.
Overall, \method’s adaptive reward margins leads to recommendations that more faithfully reflect temporally grounded human preferences.

\paragraph{Preference intensity enables discernment under temptation.}
Beyond modeling contextualized user preferences, we evaluate the model’s ability to avoid recommending items that are ultimately dispreferred, even when they appear contextually relevant.
To this end, we construct test sets from MovieLens and Steam by augmenting candidate sets with low-rated items from users’ future interactions.
While these items are rated poorly in hindsight, their later occurrence, often driven by exposure or curiosity, makes them superficially plausible as next-item recommendations, thus posing a form of contextual temptation.
To measure how well a model resists such temptations when predicting the next interaction, we define the \textit{Avoidance Rate} as the proportion of cases in which the model successfully avoids recommending them.
More metric details are provided in Appendix~\ref{appd:metrics}.

As shown in Figure~\ref{fig:exp}(b), \method consistently achieves the highest avoidance rates across benchmarks, outperforming all baselines.
These results indicate that incorporating structured feedback enables the model to internalize both positive and negative preference signals---reducing the likelihood of recommending irrelevant or disliked items while consistently adhering to user preference trajectories, and thereby enhancing overall alignment with user intent.

\paragraph{Both factors jointly enable implicit aversion modeling.}
While most preference alignment focuses on promoting desirable items, an essential aspect of human-like decision-making is the ability to deliberately avoid dispreferred options.
To evaluate this capacity, we construct a test set querying the model directly at inference time to identify the item least aligned with a user’s preferences, without providing any explicit supervision for aversion.
This setup tests whether the model’s learned preference representation implicitly encodes negative signals alongside positive ones.
We define aversion accuracy as the proportion of cases in which the model correctly identifies the least preferred item without explicit aversion supervision.

As shown in Figure~\ref{fig:exp}(c), \method consistently outperforms SFT and S-DPO with higher aversion accuracy across both datasets.
This suggests that \method internalizes a more complete structure of user preferences, capable of both attraction and avoidance.
Notably, this behavior emerges without explicit aversion labels---through alignment with structured and contextualized feedback alone, \method learns to infer items users are most likely to reject.

\paragraph{Learned patterns generalize across varying interaction contexts.}
In Figure~\ref{fig:exp}(d), we investigate \method’s robustness to variations in historical interaction lengths using the BeerAdvocate dataset. 
We partition the test set into subsets based on the number of past interactions and evaluate performance within each group.
\method exhibits sustained efficacy, consistently outperforming SFT and S-DPO with larger margins.
While all models follow similar performance trends as history length increases, \method exhibits the greatest stability, with the lowest variance in Hit Ratio@1 (8.7\% vs. 17.8\% for S-DPO). 
These results highlight \method’s adaptability to diverse context---a critical trait for real-world systems where user histories vary widely.

\begin{figure}[t]
  \centering
  \includegraphics[width=\columnwidth]{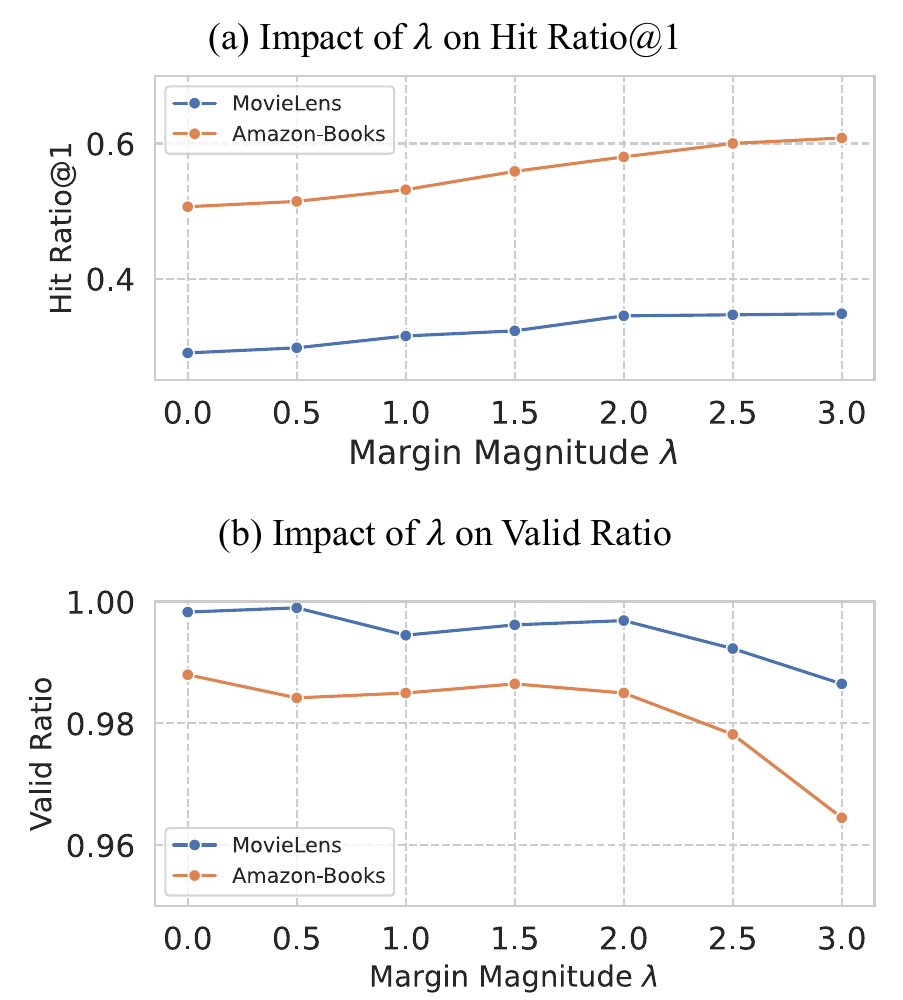}
   \caption{Sensitivity analysis of the margin parameter $\lambda$ on: (a) Hit Ratio@1 and (b) Valid Ratio across MovieLens and Amazon-Books datasets.}
  \label{fig:lambda}
\end{figure}

\section{Analysis on Margin Magnitude}

As detailed in \S~\ref{sec:method}, the parameter $\lambda$ controls the influence extent of the margin term $\gamma_r$ on preference learning. We adopt $\lambda = 2$ as the default value to balance Hit Ratio@1 (recommendation accuracy) and Valid Ratio (instruction-following capability). To further study the impact of $\lambda$ on model effectiveness, we conduct sensitivity analyses on MovieLens and Amazon-Books, with results visualized in Figure~\ref{fig:lambda}. Increasing $\lambda$ consistently elevates Hit Ratio@1, though the rate of improvement diminishes at higher values (e.g., $\lambda = 3$). However, excessively large $\lambda$ values degrade the Valid Ratio, which quantifies the model's adherence to user instructions. While Hit Ratio@1 reflects recommendation accuracy, maintaining a robust Valid Ratio ensures alignment with user intent. We recommend $\lambda \approx 2$ to harmonize both metrics.

\section{Related Work}
\textbf{Sequential recommendation} models temporal user preferences in interaction sequences, 
different from general recommendation tasks that treat user behavior as static and 
order-independent signals~\cite{rendle2009bpr,he2020lightgcn,wu2021self,ouyang2025non,ouyang2025scaled,ouyang2024improve}. 
Early methods adopt structures such as recurrent neural networks \cite{hidasi2016session}, self-attention mechanisms \cite{kang2018self}, and convolutional layers \cite{chang2021sequential} for temporal modeling.
Further methods further advance the field by incorporating graph-based structures~\cite{yu2020tagnn}, contrastive learning~\cite{xie2022contrastive,chen2022intent}, and hybrid architectures~\cite{li2020time,zhou2020s3,fan2021lighter} for improved accuracy and robustness.
Recent advances integrate LLMs for their rich semantic understanding and contextual reasoning capabilities~\cite{llara2024liao, bao2023tallrec,yuan2023go}. 

\vspace{0.05in}
\noindent
\textbf{LLM preference alignment} techniques aim to align language models' outputs with human preferences. 
Reinforcement Learning from Human Feedback (RLHF)~\cite{ouyang2022training} and DPO~\cite{rafailov2024direct} fine-tune models using pairwise preference data. 
Building on DPO, methods like IPO~\cite{azar2024general}, CPO~\cite{xu2024contrastive}, KTO~\cite{ethayarajh2024kto}, SimPO~\cite{meng2024simpo}, and ODPO~\cite{amini2024direct} further refine alignment.
Most recently, S-DPO~\cite{chen2024on} adapts alignment for sequential recommendation using list-wise negative items.
However, these methods model preferences through binary comparisons, overlooking structured preference intensity and temporal context.
More details are in Appendix~\ref{appd:related_work}.

Our work differs by  investigating what factors enable effective preference modeling in LLM-based recommendation, revealing that preference intensity and temporal context are critical yet overlooked dimensions.

\section{Conclusion}
We investigate what enables LLMs to effectively model user preferences in sequential recommendation. 
Our analysis identifies two key factors---reference intensity and recency-sensitive temporal context---that are largely collapsed by binary preference modeling.
Motivated by these findings, we propose \method, a preference optimization framework that operationalizes these factors via adaptive reward margins. Extensive experiments demonstrate that jointly modeling preference intensity and temporal context yields consistent improvements over state-of-the-art baselines.
More broadly, this work advocates a shift in LLM-based recommendation away from static, uniform preference supervision toward learning from richer interaction traces and dynamic contextual signals as first-class sources of preference information.

\section*{Acknowledgments}
This research was supported in part by the National Science Foundation under Grant No. 2452367.

\section*{Limitations}
While our results demonstrate that incorporating comprehensive and structured interaction feedback improves user preference profiling, this work adopts a simplified, sequential preference structure and considers only satisfaction delay as the contextual factor.
In reality, human decision-making reflects more complex hierarchies and richer contextual influences.
Future research should explore how to model cognitively plausible preferences across broader preference-based tasks, extending beyond recommendations.
Even within the recommendation domain, evaluations should move beyond single metrics, aiming to capture more holistic and behaviorally grounded patterns of user preference.

\bibliography{ref}

\newpage

\appendix
\section{Preliminaries}
\label{appd:prelim}
Continuing from the main paper, we outline the two-stage training paradigm that adapts existing LMs to the recommendation task, including \textit{supervised fine-tuning (SFT)} and \textit{preference alignment}.
Centering around the alignment stage, we briefly introduce direct preference optimization (DPO)~\cite{rafailov2024direct}, a technique that aligns LMs using pairwise preference data;
We then present S-DPO~\cite{chen2024on}, a recent adaptation of DPO designed specifically for sequential recommendation.


\paragraph{Supervised Fine-tuning LMs for Sequential Recommendation.}
Supervised fine-tuning~\cite{ouyang2022training, jia2025makes} (SFT) is widely adopted to adapt general-purpose LMs to recommendation tasks~\cite{llara2024liao,bao2023tallrec}.
Let $\mathbf{x}_u^t$ be the task prompt that encompasses user $u$'s interaction history $\mathcal{H}_u^t$ up to time $t$, the candidate item set $\mathcal{C}$, and other task-related descriptions.
We define $\mathbf{y}_p^t$ as the text mapping of item $i^{t+}_p \in \mathcal{C}$ that best aligns with $\mathbf{x}_u^t$'s description.
We construct the SFT training dataset $\mathcal{D}_{\text{SFT}}$ using pairwise data $(\mathbf{x}_u^t, \mathbf{y}_p^{t+}), \forall u,\forall t<N_u$, and frame the sequential recommendation as a sentence completion task.
The objective that optimizes $\pi_\theta$ is:
\begin{equation}\label{eq:sft}
    \max_\theta \quad \mathbb{E}_{(\mathbf{x}_u^t, \mathbf{y}^{t+}_p)\sim \mathcal{D}_{\text{SFT}}} \left[ \log \pi_\theta(\mathbf{y}^{t^{+}}_p|\mathbf{x}_u^t)  \right].
\end{equation}
The LM fine-tuned with this objective on $\mathcal{D}_{\text{SFT}}$ is denoted as $\pi_{\text{SFT}}$. For brevity, we omit the timestamp signs in all subsequent equations unless its inclusion is essential for clarity.

\paragraph{Aligning LLM with Human Preference Feedback.}
While optimizing the SFT objective effectively adapts LMs to the downstream task, recent studies indicate that models still struggle to align outputs with human judgments of quality~\cite{ziegler2019fine,stiennon2020learning,rafailov2024direct, zhang2025pretrained}.
To address this, a reward model $r(\mathbf{x}, \mathbf{y})$ is introduced to estimate output quality assessed by humans, aiming to maximize the expected reward.

To train the reward model, a dataset of comparisons $D = \{\mathbf{x}^{(i)}, \mathbf{y}_w^{(i)}, \mathbf{y}_l^{(i)}\}_{i=1}^{N}$ is constructed, where $\mathbf{y}_w^{(i)}$ and $\mathbf{y}_l^{(i)}$ denotes the preferred and dispreferred output generated based on $\mathbf{x}^{(i)}$, respectively.
The alignment objective with the learned reward function is then defined as:
\begin{align}\label{eq:reward}
\max_{\theta} \quad & \mathbb{E}_{x \sim \mathcal{D}, \mathbf{y} \sim \pi_\theta(\cdot | \mathbf{x})} \Bigl( \left[ r(\mathbf{x}, \mathbf{y}) \right] \notag \Bigr. \\
& \Bigl. - \beta D_{\text{KL}} \left[ \pi_\theta(\mathbf{y} | \mathbf{x}) \| \pi_{\text{ref}}(\mathbf{y} | \mathbf{x}) \right] \Bigr),
\end{align}
where $\beta$ is the parameter controlling the deviation from the reference model $\pi_{\text{ref}}$, and $\pi_{\text{SFT}}$ is commonly used as the reference model.
Based on Equation~\ref{eq:reward}, a recent work DPO~\cite{rafailov2024direct}, employs the Bradley-Terry~\cite{bradley1952rank} (BT), $P\left(\mathbf{y}_w \succ \mathbf{y}_l \mid \mathbf{x}\right)=\sigma\left(r\left(\mathbf{x}, \mathbf{y}_w\right)-r\left(\mathbf{x}, \mathbf{y}_l\right)\right)$, to express the probability of human preference data in terms of the optimal policy rather than the reward model, they derive the objective based on pairwise preference data as:
\begin{align}\label{eq:dpo}
    \min_\theta & -\mathbb{E}_{(\mathbf{x}, \mathbf{y}_w, \mathbf{y}_l) \sim D} \Bigl[ \log \sigma \Bigl( \beta \log \frac{\pi_\theta (\mathbf{y}_w | \mathbf{x})}{\pi_{\text{ref}}(\mathbf{y}_w | \mathbf{x})} \notag\\
    & - \beta \log \frac{\pi_\theta (\mathbf{y}_l | \mathbf{x})}{\pi_{\text{ref}}(\mathbf{y}_l | \mathbf{x})} \Bigr) \Bigr].
\end{align}

The above preference modeling paradigm aligns naturally with recommendation tasks, with both being preference-based decision-making. 
Building upon DPO, a recent effort named S-DPO~\cite{chen2024on} is proposed to further align LLM-based recommenders to user preference.
They propose to pair each positive item with multiple negative items generated by random sampling as preference data, and revise the alignment objective as:

{
\fontsize{10.5pt}{12pt}\selectfont
\begin{align}\label{eq:s-dpo}
    \min_\theta &-\mathbb{E}_{(\mathbf{x}, \mathbf{y}_w, \mathcal{T}_d) \sim D} \biggl[ \log \sigma \biggl( -\log \sum_{\mathbf{y}_d \in \mathcal{T}_d} \exp \Bigl( \notag \\
     & \beta \log \frac{\pi_\theta (\mathbf{y}_d | \mathbf{x})}{\pi_{\text{ref}}(\mathbf{y}_d | \mathbf{x})} - \beta \log \frac{\pi_\theta (\mathbf{y}_p | \mathbf{x})}{\pi_{\text{ref}}(\mathbf{y}_p | \mathbf{x})} \Bigr) \biggr) \biggr],
\end{align}
}

\noindent
where $\mathcal{T}_d$ contains the item titles of multiple dispreferred items\footnote{We use positive/negative, as well as preferred/dispreferred interchangeably in the following content.}.

\section{Derivation of Preference Distribution}
\label{appd:derivation}

In the standard Bradley-Terry model, the probability that candidate $i$ beats candidates $j$ is 
\begin{equation}
\begin{aligned}
P&(\mathbf{y}_i \succ \mathbf{y}_j) = \sigma\left(r\left(\mathbf{x}, \mathbf{y}_w\right)-r\left(\mathbf{x}, \mathbf{y}_l\right)\right) \\
& = \frac{\exp\left(r\left(\mathbf{x}_u, \mathbf{y}_i\right)\right)}{\exp\left(r\left(\mathbf{x}_u, \mathbf{y}_i\right)\right)+\exp\left(r\left(\mathbf{x}_u, \mathbf{y}_i\right)\right)},
\end{aligned}
\end{equation}
where $r(\cdot)$ is the reward model.
We will only use $w_i$ to represent the candidate-specific probability $\exp\left(r\left(\mathbf{x}_u, \mathbf{y}_i\right)\right)$ in subsequent equations for brevity. Now suppose we wish to include a margin term $\gamma_{ij}$, then the pairwise probability is defined as 
\begin{equation}
\begin{aligned}
 P(\mathbf{y}_i \succ \mathbf{y}_j) =  \frac{w_i \exp \left( -\gamma_{ij} \right)}{w_i \exp \left(-\gamma_{ij} \right) + w_j} \\
\end{aligned}
\end{equation}
where we assume $\gamma_{ij} = -\gamma_{ji}$. Specifically, we can use the Plackett-Luce model decomposes a ranking $i_1 \succ i_2 \succ i_k \succ \cdots \succ i_K$ into sequential choices competition. 
Therefore, at each step $t$, the wining (got selected) probability $i_k$ is proportional to its weight, i.e., $w_k=\exp\left(r\left(\mathbf{x}_u, \mathbf{y}_k\right)\right)$.
Now the added margin term $\gamma_{ij}$ modifies the competition by giving each candidate an extra boost (or penalty) when facing an opponent. In other words, when candidate $i$ competes against candidate $j$ (within the remaining set) its effective strength is boosted by the factor $\exp(-\gamma_{ij})$. 
Then, by an extension of Luce's choice axiom, we can get the probability of choosing candidate $i$ from the set $\mathcal{C}$ is proportional to its effective weight:
\begin{equation}
\begin{aligned}
P(i & \text { chosen from } \mathcal{C})= \\
& \frac{w_{i} \exp \left(-\sum_{j \in \mathcal{C} \backslash\{i\}} \gamma_{i j}\right)}{\sum_{k \in \mathcal{C}} w_{k} \exp \left(-\sum_{j \in \mathcal{C} \backslash\{k\}} \gamma_{k j}\right)} .
\end{aligned}
\end{equation}
Let $\tau=(\tau(1), \tau(2), \ldots, \tau(K))$ be a full ranking of $K$ candidates. We construct the ranking sequentially. At step $r$, let 
\begin{equation}
\mathcal{C}_r=\mathcal{C} \backslash\{\tau(1), \tau(2), \ldots, \tau(r-1)\}
\end{equation}
be the remaining set. Then the probability that candidate $\tau(r)$ is selected at step $r$ will be,
\begin{equation}
\begin{aligned}
P & (\tau(r)  \mid \tau(1), \ldots, \tau(r-1))= \\ 
&\frac{w_{\tau{(r)}} \exp \left(-\sum_{j \in \mathcal{C}_r \backslash\{\tau(r)\}} \gamma_{\tau(r) j}\right)}{\sum_{k \in \mathcal{C}_r} w_{\tau{(k)}} \exp \left(-\sum_{j \in \mathcal{C}_r \backslash\{k\}} \gamma_{k j}\right)} .
\end{aligned}
\end{equation}
We can thereby get the likelihood of the full ranking by the chain rule, 
\begin{equation}
\begin{aligned}
&  P ( \tau \mid \mathcal{C})= \\
& \prod_{r=1}^{K-1} \frac{  w_{\tau{(r)}} \exp \left(-\sum_{j \in \mathcal{C}_r \backslash\{\tau(r)\}} \gamma_{\tau(r) j}\right)}{\sum_{k \in \mathcal{C}_r} w_{\tau_{(k)}} \exp \left(-\sum_{j \in \mathcal{C}_r \backslash\{k\}} \gamma_{k j}\right)}
\end{aligned}
\end{equation}
In the recommendation setting we are especially interested in penalizing the positive item’s “win” relative to each negative, which means one might only apply a margin from the positive item to each negative. 
Therefore, we can derive the preference distribution of recommendation case given interactions $\mathbf{x}_u$ of user $u$, multiple negative items $\mathbf{y}_d \in \mathcal{T}_d$ and the positive item $\mathbf{y}_p$:
\begin{equation}
\begin{aligned}
P ( & \mathbf{y}_p \succ \mathbf{y}_d, \forall \mathbf{y}_d \in \mathcal{T}_d  \mid \mathbf{x}_u, \mathbf{y}_p, \mathcal{T}_d)  = \\
& \frac{w_p \exp \left(-\sum_{j=1}^{K-1} \gamma_{p, d_j}\right)}{w_p \exp \left(-\sum_{j=1}^{K-1} \gamma_{p, d_j}\right)+\sum_{j=1}^{K-1} w_{d_j}} .
\end{aligned}
\end{equation}
Notably, the ranking likelihood would reduce to the standard Plackett–Luce model if the margin term $\gamma=0$ for all pairs.

\section{Prompt Examples}\label{appd:prompt}
\begin{figure}[t]
  \centering
  \includegraphics[width=\linewidth]{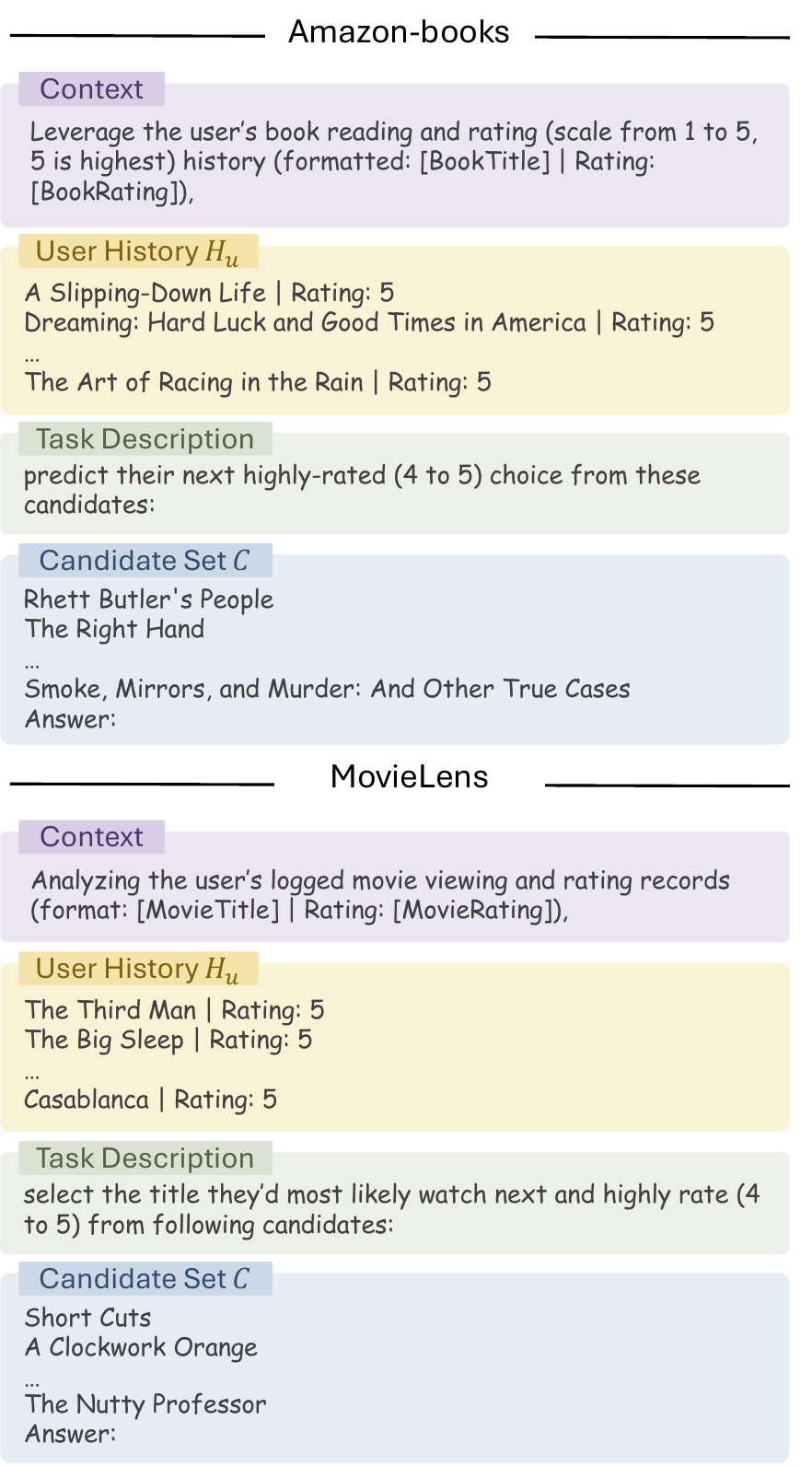}
   \caption{Textual prompt examples for Amazon-books and MovieLens.}
  \label{fig:prompt}
\end{figure}

We refer the prompts used in previous works~\cite{chen2024on, llara2024liao} to construct prompts utilized in our work.
Examples in Figure~\ref{fig:prompt} demonstrates the prompts for sequential recommendation.

\section{Related Work}\label{appd:related_work}
\paragraph{Sequential Recommendation.}
Sequential recommendation aims to model user preferences by capturing temporal patterns in interaction sequences. Early approaches, such as GRU4Rec~\cite{hidasi2016session}, leveraged recurrent neural networks (RNNs) to encode sequential dependencies, while SASRec~\cite{kang2018self} introduced self-attention mechanisms to better capture long-range dependencies. 
Convolutional-based methods like Caser~\cite{chang2021sequential} explored local patterns in sequences using convolutional filters. 
Recent state-of-the-art methods have further advanced the field by incorporating graph-based structures~\cite{yu2020tagnn}, contrastive learning~\cite{xie2022contrastive,chen2022intent}, and hybrid architectures~\cite{li2020time,zhou2020s3,fan2021lighter} for improved accuracy and robustness.

\paragraph{LLMs for Recommendation.}
The integration of LLMs into sequential recommendation has gained momentum due to their ability to leverage rich semantic knowledge and contextual understanding. 
LLMs are typically integrated by encoding item descriptions, user reviews, or interaction histories as textual inputs, enabling the model to capture nuanced item characteristics and user preferences. 
For instance, LLaRA~\cite{llara2024liao} employs classical sequential recommender systems to generate item embeddings, which are then fused with sequential interaction data to improve recommendation accuracy.
TALLRec~\cite{bao2023tallrec} fine-tunes LLMs on user-item interaction sequences, treating recommendations as a text generation task to predict the next item. 
Other approaches tackles the task from prompting~\cite{geng2022recommendation,gao2023chat,lyu2023llm} or multi-modal data exploitation~\cite{yuan2023go}. 

Recent work has also begun to incorporate multiple and fine-grained preference signals into LLM-based recommenders.
In particular, ELCoRec~\cite{chen2024elcorec} integrates numerical and categorical features (e.g., ratings and temporal attributes) with textual representations through feature co-propagation, enhancing the expressiveness of input representations.
However, these methods primarily operate at the input or representation level, relying on the model to implicitly infer the relative importance of heterogeneous preference signals during training.

In contrast, our work focuses on preference alignment, explicitly modeling how strongly different preference signals should influence learning by modulating the alignment objective itself.
Rather than encoding fine-grained or temporal signals solely as features, we incorporate preference intensity and recency-sensitive temporal context directly into adaptive reward margins, enabling principled control over preference learning dynamics beyond representation enrichment.

\paragraph{LLM Alignment.}
LLM alignment techniques aim to align general-purpose LMs' outputs with human preferences, ensuring that generated content is both useful and safe. 
While not specifically designed for recommendation tasks, these methods have inspired advancements in preference modeling. 
Early approaches like Reinforcement Learning from Human Feedback (RLHF)~\cite{ouyang2022training, schulman2017proximal} laid the foundation by using reinforcement learning to fine-tune models based on human feedback. 
DPO~\cite{rafailov2024direct} emerged as a simpler and more efficient alternative, directly optimizing preference data without requiring explicit reward modeling. 
Building on DPO, methods like IPO~\cite{azar2024general}, CPO~\cite{xu2024contrastive}, KTO~\cite{ethayarajh2024kto}, SimPO~\cite{meng2024simpo}, and ODPO~\cite{amini2024direct} further refine alignment by addressing limitations such as capturing fine-grained preference hierarchies, reducing reward hacking, improving robustness to noisy feedback, and enhancing generalization across diverse user contexts.
Most recently, S-DPO~\cite{chen2024on} adapts alignment techniques specifically for recommendation tasks, focusing on sequential user preferences and improving the personalization of LLM-based recommenders.

Beyond recommendation, preference alignment has been explored across diverse domains, 
including logical reasoning~\cite{yuan2025superficial, yuan2025mitigating, yuan2026behavior, ma2026fipo}, audio-language reasoning~\cite{diao2025soundmind} and multimodality~\cite{zhang2025overcoming, zhang2025knowing, diao2026addressing}, demonstrating its broad applicability as a 
general framework for aligning model behavior with human preferences.

\section{Experimental Settings}

\subsection{Datasets}
\label{appd:dataset}

\begin{table}[t]
\begin{adjustbox}{width=1.\linewidth,center}
\begin{tabular}{l|c|c|c}
\toprule
Dataset           & \# Sequence  & \# Items & \# Interactions
\\ \midrule
MovieLens               &   6,040        &     3,952        &           994,169       \\
Amazon-Books             &   5,103        &   38,203      &  62,290             \\
Steam  &   3,171      &    4,251       &  82,072    \\
BeerAdvocate &     4,724      &    6,105         &     91,207 \\
LastFM & 982 & 107,296 & 307,829
\\ \bottomrule   
\end{tabular}
\end{adjustbox}
\caption{Statistics of datasets}
\label{tab:dataset}
\end{table}

We use five widely used real-world sequential recommendation datasets for evaluation, including 
\textit{MovieLens-1M}\footnote{\url{https://grouplens.org/datasets/movielens/1m/}}~\cite{harper2015movielens}, 
\textit{Amazon-books}\footnote{\url{https://nijianmo.github.io/amazon/index.html}}~\cite{ni2019justifying}, 
\textit{Steam}\footnote{\url{https://github.com/kang205/SASRec}}~\cite{kang2018self}, 
\textit{BeerAdvocate}\footnote{\url{https://cseweb.ucsd.edu/~jmcauley/datasets.html\#multi_aspect}}~\cite{leskovec2012learning}, and \textit{LastFM}\footnote{\url{http://ocelma.net/MusicRecommendationDataset/lastfm-1K.html}}~\cite{celma2010music}.
We demonstrate the dataset statistics in Table~\ref{tab:dataset}. 
The MovieLens-1M dataset is sourced from the MovieLens platform and contains 1 million ratings from 6,000 users on 4,000 movies. The Amazon-Books dataset is a subset of the Amazon Review dataset and comprises 22 million user interactions, reviews, and ratings for 2 million books from 8 million users. The Steam dataset includes user interactions with games—such as purchases, playtime, and reviews—from the Steam platform. The BeerAdvocate dataset collects beer reviews that cover multiple sensory aspects along with overall ratings.
The LastFM dataset comprises detailed music listening records for nearly 1,000 users, including user profiles with demographic information, artist and track identifiers, and precise timestamps for each listening event.

For each dataset, we filter out items and users with fewer than 20 interactions. To prevent information leakage during training and evaluation, we adopt the leave-last-two splitting method to divide the datasets into training, validation, and test sets. We build a candidate set of 20 items for each user sequence, from which the model selects the next item. During training, this set comprises 10 subsequent interactions (ensuring that the correct item is always included) and 10 randomly sampled non-interacted items. For validation and testing, the candidate set consists of the correct item plus 19 randomly sampled non-interacted items.
To align with the task objective of recommending the most likely favorable item as the next interaction, we follow classical sequential recommendation settings by considering only highly rated items (ratings 4 to 5 on a scale of 1 to 5) from subsequent interactions as the positive item (i.e., the correct answer)~\cite{li2024calrec}. The same process is applied to the validation and test sets; we only retain user sequences whose next item is highly rated. Meanwhile, we preserve all historical interactions and their corresponding ratings in the user history sequence for comprehensive user preference profiling.

For Steam and LastFM, since they lack explicit rating signals, we convert play-hours and play-count respectively to a 1-to-5 scale structured rating based on its percentile ranking. 
For example, if a user's playtime for a game falls within the top 20\% compared to other players, the corresponding user-item pair is assigned a rating of 5.

\subsection{Baselines}
\label{appd:baselines}
We include the following baseline models for performance comparison:
\begin{itemize}
    \item GRU4Rec~\cite{hidasi2016session} is a recurrent neural network-based model that captures sequential patterns in user interaction sequences session-based recommendation.
    \item Caser~\cite{tang2018personalized} is a convolutional neural network-based model that learns both local and sequential patterns in user-item interactions using convolutional filters. 
    \item SASRec~\cite{kang2018self} is a transformer-based model that leverages self-attention to capture long-range dependencies and dynamic user preferences in sequential recommendation.
    \item LLaMA-3~\cite{dubey2024llama} is a general-purpose LLM with strong semantic reasoning capabilities. We adapt it to sequential recommendation by treating it as a text prediction problem.
    \item Qwen2.5~\cite{bai2023qwen} is a recent LLM developed by Alibaba, optimized for instruction-following and multi-turn dialogue tasks.
    \item DPO~\cite{rafailov2024direct} is a preference alignment technique that fine-tunes models using pairwise preference data. In this work, we construct preference data based on explicit preference feedback.
    \item SimPO~\cite{meng2024simpo} is an extension of DPO that directly optimizes pairwise preferences without requiring explicit reward models or complex sampling strategies for improved efficiency and scalability.
    \item S-DPO~\cite{chen2024on} is a variant of DPO specifically adapted for sequential recommendation that incorporates list-wise negative items in preference alignment.
\end{itemize}


\subsection{Implementation Details}
\label{appd:implementation}

All experiments were conducted on a maximum of 8 NVIDIA RTX A6000 GPUs, each with 48GB of VRAM. Our framework is implemented using Python 3.10.6, PyTorch 2.2.2, and Huggingface Transformers 4.43.3. For all LLM-based recommenders, we employ LLaMA 3.1 8B~\cite{dubey2024llama} and Qwen2.5-7B~\cite{bai2023qwen} as the base models for both SFT and alignment. 
During training, we set the learning rate to 1e-5 for all LLM-based recommenders and use the AdamW optimizer. Additionally, we apply a 5\% warm-up strategy and adjust the learning rate using a cosine scheduler. A global batch size of 128 is used to balance training efficiency and memory consumption. The maximum sequence length is tailored to each dataset based on the features involved and the average title lengths. 
We set $\beta=1$ for all preference optimization approaches. For multi-negative preference learning, including S-DPO and our proposed \method, we adopt the S-DPO settings and fix the number of negatives at 3.
In particular, we set the margin term in SimPO as 2 and set the parameter $\lambda$ in our method as 2. 
Finally, following the prompt format provided in Appendix~\ref{appd:prompt}, we create several additional prompt templates and randomly sample one for each user sequence during training and evaluation to ensure model flexibility and generality.
For all traditional recommenders, we follow the settings from previous work~\cite{chen2024on} by setting the learning rate to 0.001, the batch size to 256, and using the Adam optimizer for model optimization. 

\subsection{Evaluation Metrics}
\label{appd:metrics}

As mentioned in \S~\ref{sec:setup}, we primarily employ two metrics to evaluate model effectiveness: Hit Ratio@1, which measures how accurately the model recommends the correct item, and Valid Ratio, which assesses whether the model follows instructions to generate outputs in the required format. In \S~\ref{sec:exp_analysis}, we introduce two additional metrics—\textbf{\textit{Adherence Rate}} and \textbf{\textit{Avoidance Rate}}—both derived from Hit Ratio@1. These metrics evaluate the model's ability to adhere to contextualized user preferences and avoid recommending unfavorable (unsatisfactory) items for the next interaction, with higher values indicating better performance.

In our main experiment, the candidate sets during testing include the last item from the user's full sequence, typically a highly rated item (rating 4 to 5 on a scale of 1 to 5), with the remaining candidates randomly sampled from the non-interacted set. 
Note that we use rating to denote the preference hierarchy, yet it can be derived from either implicit or explicit feedback.
\textbf{In the contextualized preference adherence experiment, the candidate set for testing includes at least two highly-rated items from the subsequent sequence}. We follow the rule described in \S~\ref{sec: preliminary} to designate the positive item as the one with the smallest time latency $\Delta_t$ relative to the prediction timestamp $t$. A high \textbf{\textit{Adherence Rate}} indicates that the model consistently recommends the positive item among all highly-rated candidates.

For the unfavorable item avoidance experiment, we construct the test set by selecting user sequences where the last interaction is low-rated (rating 1 to 2). Instead of measuring whether the model recommends this low-rated item, we assess whether it favors the randomly sampled candidates over the unfavorable item. Thus, a high \textbf{\textit{Avoidance Rate}} signifies that the model successfully avoids recommending unfavorable items to users.



\section{Discussion on Potential Risks}
This work focuses on methodological limitations and performance gaps, and does not touch ethical, societal, or deployment-related risks such as user manipulation, fairness, or privacy concerns.

\section{Discussion on Training and Deployment Cost}
Our experiments use an 8B-parameter backbone with standard parameter-efficient fine-tuning, which is well within the scale already deployed in many production recommendation stacks; training is a one-time offline cost, while inference reuses the same model for diverse tasks. 
In addition, \method is \textbf{orthogonal} to model size, architecture, and optimizer~\cite{pang2026htmuon}: the objective can be applied to smaller or specialized backbones, or combined with distillation/compression, so it does not inherently require heavier deployment than existing LLM-based recommenders.

\end{document}